\documentclass[range]{ar2e}

\usepackage{amsmath,amssymb,bm}
\usepackage{epsfig,graphicx,psfrag}
\def\be{\begin{equation}}
\def\ee{\end{equation}}
\def\ba{\begin{eqnarray}}
\def\ea{\end{eqnarray}}

\begin{document}



\newcommand {\tbf}[1] {\textbf{#1}}
\newcommand {\tit}[1] {\textit{#1}}
\newcommand {\tmd}[1] {\textmd{#1}}
\newcommand {\trm}[1] {\textrm{#1}}
\newcommand {\tsc}[1] {\textsc{#1}}
\newcommand {\tsf}[1] {\textsf{#1}}
\newcommand {\tsl}[1] {\textsl{#1}}
\newcommand {\ttt}[1] {\texttt{#1}}
\newcommand {\tup}[1] {\textup{#1}}

\newcommand {\figwidth} {100mm}
\newcommand {\Ref}[1] {Reference~\cite{#1}}
\newcommand {\Sec}[1] {Section~\ref{#1}}
\newcommand {\Eq}[1] {Eq.~\ref{#1}}
\newcommand {\App}[1] {Appendix~\ref{#1}}
\newcommand {\Chap}[1] {Chapter~\ref{#1}}
\newcommand {\Fig}[1] {Figure~\ref{#1}}
\newcommand {\bul} {$\bullet$ }   
\newcommand {\fig}[1] {Figure~\ref{#1}}   
\newcommand {\imp} {$\Rightarrow$}   
\newcommand {\impt} {$\Rightarrow$}   
\newcommand {\impm} {\Rightarrow}   
\newcommand {\vect}[1] {\mathbf{#1}}
\newcommand {\hvect}[1] {\hat{\mathbf{#1}}}
\newcommand {\del} {\partial}
\newcommand {\eqn}[1] {Equation~(\ref{#1})}
\newcommand {\tab}[1] {Table~\ref{#1}} 
\newcommand {\ten}[1] {\times10^{#1}}
\newcommand {\Bra}[2] {\mbox{}_{#2}\left.\left\langle #1 \right.\right|}
\newcommand {\Ket}[2] {\left.\left| #1 \right.\right\rangle_{#2}}
\newcommand {\im} {\mathrm{Im}}
\newcommand {\re} {\mathrm{Re}}
\newcommand {\braket}[4] {\mbox{}_{#3}\langle #1 | #2 \rangle_{#4}}
\newcommand {\dotprod}[4] {\mbox{}_{#3}\langle #1 | #2 \rangle_{#4}}
\newcommand {\trace}[1] {\text{tr}\left(#1\right)}

\newcommand{\beq}{\begin{equation}}
\newcommand{\eeq}{\end{equation}}

\newcommand{\natc}{N@C$_{60}$}
\newcommand{\natcseventy}{N@C$_{70}$}
\newcommand{\heatc}{He@C$_{60}$}
\newcommand{\heatcseventy}{He@C$_{70}$}
\newcommand{\patc}{P@C$_{60}$}

\newcommand{\csixty}{C$_{60}$}
\newcommand{\cseventy}{C$_{70}$}
\newcommand{\ceightyfour}{C$_{84}$}
\newcommand{\ceightytwo}{C$_{82}$}
\newcommand{\ceighty}{C$_{80}$}
\newcommand{\cseventyeight}{C$_{78}$}
\newcommand{\cseventysix}{C$_{76}$}
\newcommand{\mus}{$\rm{\mu}$s}
\newcommand{\cstwo}{CS$_2$}
\newcommand{\stwocltwo}{S$_2$Cl$_2$}

\newcommand{\nfourteen}{$^{14}$N}
\newcommand{\pthirtyone}{$^{31}$P}
\newcommand{\ptone}{$^{31}$P}
\newcommand{\nfifteen}{$^{15}$N}
\newcommand{\cthirteen}{$^{13}$C}
\newcommand{\sitwonine}{$^{29}$Si}
\newcommand{\sitwoeight}{$^{28}$Si}
\newcommand{\biiso}{$^{209}$Bi}

\newcommand{\beqa}{\begin{eqnarray}}
\newcommand{\eeqa}{\end{eqnarray}}
\newcommand{\w}{\omega}
\newcommand{\thetai}{\theta_i}
\newcommand{\raiz}{\mbox{$\textstyle \frac{1}{\sqrt{2}}$}}
\newcommand{\ket}[1]{\left| #1 \right\rangle}
\newcommand{\bra}[1]{\left\langle #1 \right|}
\newcommand{\proj}[1]{\ket{#1}\bra{#1}}
\newcommand{\av}[1]{\langle #1\rangle}
\newcommand{\inprod}[2]{\braket{#1}{#2}}
\newcommand{\upket}{\ket{\uparrow}}
\newcommand{\downket}{\ket{\downarrow}}

\newcommand{\ACIE}{Ang. Chem. Int. Ed.}
\newcommand{\APL}{Appl. Phys. Lett.}
\newcommand{\CCOMM}{Chem. Comm.}
\newcommand{\CPL}{Chem. Phys. Lett.}
\newcommand{\IJTP}{Int. J. Thermophys.}
\newcommand{\JCP}{J. Chem. Phys.}
\newcommand{\JAP}{J. Appl.. Phys.}
\newcommand{\JACS}{J. Am. Chem. Soc.}
\newcommand{\JMOLLIQ}{J. Mol. Liq.}
\newcommand{\JMR}{J. Mag. Res.}
\newcommand{\JPC}{J. Phys. Chem.}
\newcommand{\JPCM}{J. Phys. Cond. Matt.}
\newcommand{\MOLPHYS}{Mol. Phys.}
\newcommand{\PCCP}{Phys. Chem. Chem. Phys.}
\newcommand{\PTRSA}{Phil.~Trans.~R.~Soc. A}
\newcommand{\PR}{Phys. Rev.}
\newcommand{\PRA}{Phys. Rev. A}
\newcommand{\PRB}{Phys. Rev. B}
\newcommand{\PRL}{Phys. Rev. Lett.}
\newcommand{\PNAS}{Proc. Natl. Acad. Sci.}
\newcommand{\RSI}{Rev. Sci. Inst.}

\newcommand{\tr}{$t_{\rm{r}}$}
\newcommand{\ttwo}{$T_2$}
\newcommand{\ttwoi}{T$_{2,i}$}
\newcommand{\ttwoo}{T$_{2,o}$}
\newcommand{\tonei}{T$_{1,i}$}
\newcommand{\toneo}{T$_{1,o}$}
\newcommand{\tone}{$T_1$}
\newcommand{\tonen}{$T_{\rm{1n}}$}
\newcommand{\tonee}{$T_{\rm{1e}}$}
\newcommand{\ttwon}{$T_{\rm{2n}}$}
\newcommand{\ttwoe}{$T_{\rm{2e}}$}
\newcommand{\ttwoq}{\rm{T}_2}
\newcommand{\toneq}{\rm{T}_1}
\newcommand{\ttwoiq}{\rm{T}_{2,\emph{i}}}
\newcommand{\ttwooq}{\rm{T}_{2,\emph{o}}}
\newcommand{\toneiq}{\rm{T}_{1,\emph{i}}}
\newcommand{\toneoq}{\rm{T}_{1,\emph{o}}}

\newcommand{\tsd}{$T_{\rm SD}$}

\newcommand{\natsi}{$^{\rm nat}$Si}



\title{Hybrid solid state qubits: the powerful role of electron spins}


\author{John J. L. Morton$^{1,2}$ and 
Brendon W. Lovett$^{1,3}$
\affiliation{$^1$Dept.\ of Materials, University of Oxford, Parks Rd, Oxford, OX1 3PH, UK\\
$^2$Clarendon Laboratory, University of Oxford, Parks Rd, Oxford, OX1 3PU, UK\\
$^3$Department of Physics, Heriot Watt University, Edinburgh, EH14 4AS}
}


\begin{abstract}
We review progress on the use of electron spins to store and process quantum information, with particular focus on the ability of the electron spin to interact with multiple quantum degrees of freedom. We examine the benefits of hybrid quantum bits (qubits) in the solid state that are based on coupling electron spins to nuclear spin, electron charge, optical photons, and superconducting qubits. These benefits include the coherent storage of qubits for times exceeding seconds,  fast qubit manipulation, single qubit measurement, and scalable methods for entangling spatially separated matter-based qubits. In this way, the key strengths of different physical qubit implementations are brought together, 
laying the foundation for practical solid-state quantum technologies.
\end{abstract}

\maketitle

\section*{Glossary}
{\bf ESR/EPR:} Electron spin/paramagnetic resonance \\
{\bf ENDOR:} Electron nuclear double resonance \\
{\bf NMR:} Nuclear magnetic resonance \\
{\bf QIP:} Quantum information processing \\
{\bf 2DEG:} Two-dimensional electron gas \\
{\bf MOS:} Metal oxide semiconductor \\ 
{\bf FET:} Field effect transistor\\ 
{\bf SET:} Single-electron transistor \\ 
{\bf QPC:} Quantum point contact: a narrow constriction between two conducting regions whose conductance is extremely sensitive to nearby charge.\\ 
{\bf Qubit:} A quantum bit, i.e. a two level quantum system that is the basic building block of a quantum computer.\\ 
{\bf Quantum computer:} A register of interacting qubits on which unitary operations (logic gates) are performed to execute quantum algorithms. \\ 
{\bf QD:} Quantum dot: a semiconductor structure in which electrons can be confined, in all three dimensions, on a nanometer length scale. \\ 
{\bf Entanglement:} exists in two or more qubits that cannot be described as a product of individual qubit states. Leads to non-classical correlations in qubit measurements.\\ 
{\bf Magneto-optical Kerr Effect (MOKE):} gives a material a polarization-dependent refractive index, and used to measure single spins through the rotation of linearly polarized light. \\ 
{\bf ISC:} Intersystem crossing: an electronic transition between a singlet and a triplet. It is forbidden in lowest order. \\ 
{\bf Trion:} A composite particle consisting of two electrons and one hole or two holes and one electron.\\ 
{\bf Decoherence:} The loss of phase information in a quantum state or qubit, characterised by the decay time constant \ttwo.\\ 
{\bf Hole:} A quasi particle generated when an electron is removed from the valence band. Heavy holes are those close to the band edge with large effective mass\\ 

\section{Introduction}
\label{sec:intro}

If the full potential of quantum information processing can be realized, its implications will be far reaching across a range of disciplines and technologies. The extent of these implications is certainly not yet fully known, but the revolutionary effect that quantum physics would have on cryptography, communication, simulation and metrology is well established~\cite{nielsen00, kok10}.

Even though building a quantum computer is very challenging, research focussed on achieving this goal can already claim to have produced some of the most high impact scientific results of the last decade. One of the reasons for this is that the language of quantum information has shown how to bring together and find new links between work on a staggering variety of physical systems. 

The past fifteen years has seen candidate embodiments of quantum bits tested to their limits of coherence time, and in some cases control over small number of such systems has become refined enough to permit the demonstration of basic quantum logic gates. However, there has been an increasing awareness that the challenge of faithfully storing, processing and measuring quantum information within physical systems is sufficiently great so as to discourage relying on one quantum degree of freedom alone.

Furthermore, classical information processors use different physical systems to encode information at different stages of their operation: for example, charge states in semiconductors for processing information, the orientation of magnetic domains in hard disks for longer term storage, and optical photons for transmitting data. 

\begin{figure}[t]
{\includegraphics[width=\columnwidth]{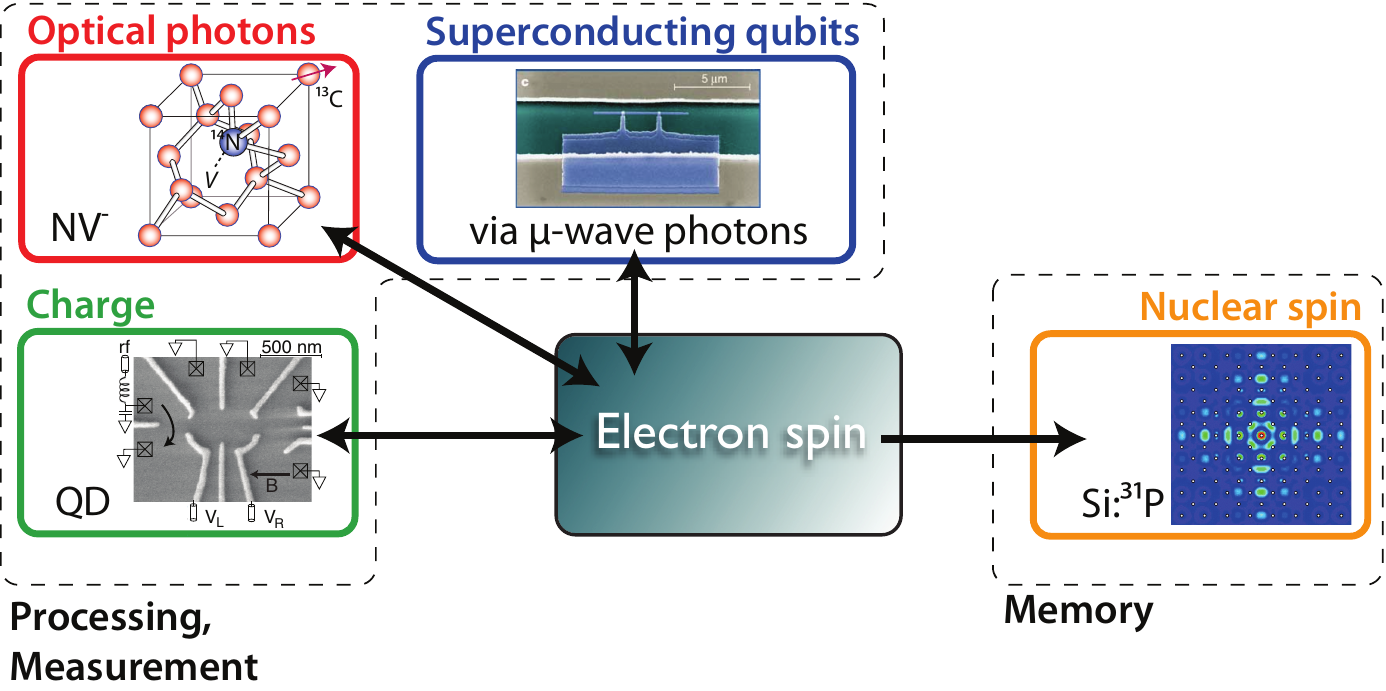}}
\caption{Electron spins in the solid state interact with several other degrees of freedom, including nuclear spins, optical photons, charge and, potentially, superconducting qubits via circuit quantum electrodynamics. In each case, example systems are illustrated. Such interactions can be harnessed to enhance the processing, coherent storage and measurement of quantum information. [Figure includes extracts from Refs~\cite{kok10} and \cite{wallraff04} with permission from Macmillam Publishers Ltd; \cite{barthel09}, Copyright (2009) by the American Physical Society http://prl.abstract.org/PRL/v103/i16/e160503; and \cite{koiller05a} with permissions from Anais da Academia Brasileira de Ci\^encias.]}
\label{summary1}
\end{figure}

For a quantum computer to benefit from such an optimised use of resources, it must be able to transfer quantum information coherently between different degrees of freedom. Within the solid state, the electron spin exhibits a number of interactions that could be harnessed for this hybridising of quantum information: with nuclear spins which benefit from long coherence times, or with charge states, or optical states either of which could be used to measure, manipulate, or even entangle electron spins (see Fig.~\ref{summary1}). 

In this review we will examine the ways in which electron spin qubits may couple to different degrees of freedom in the solid state and how this is being used to hybridise quantum information. We will distinguish between two regimes of coupling:

\begin{enumerate}
\item A weak interaction that provides an opportunity to transfer some (small) amount of classical information between the electron spin and another degree of freedom. This kind of interaction can be very important in few-to-single spin measurement, for example a small change in conductivity or photoluminescence depending on the state of an electron spin.
\item A stronger interaction capable of providing the coherent transfer of a quantum state between two degrees of freedom, with sufficient fidelity to permit the storage of quantum information or performing entangling operations.
\end{enumerate}

\section{Electron spin quantum bits}
\label{espins}

In order to understand how to use electron spins linked with other degrees of freedom in hybrid quantum processor architectures, we must first understand the different forms that electron spin qubits can take. In this section, we examine some important implementations of electron spin qubits, with a particular focus on those in the solid state.

\subsection{Quantum dots: artificial atoms}
\label{subsec:qds}

Quantum dots (QDs) are artificial structures that are fabricated so that electronic states are confined, on a nanometer length scale, in all three spatial dimensions. They are typically divided into two broad classes. First, there are lithographically defined structures consisting of a quantum well semiconductor heterostructure that confines a two dimensional electron gas (2DEG). The other dimensions are defined by lithographically deposited electrical top gates, (for an example configuration, see Fig.~\ref{summary1}). This allows one, two or more dots to be deposited side-by-side; gate electrodes are used to control the number of charges within the structure, and allow a single electron spin to be isolated. 

\begin{figure}[t]
  \begin{psfrags}
  {\scriptsize
   \psfrag{u}{$\sigma^+$}
   \psfrag{v}{$\sigma^-$}
   \psfrag{a}{$\ket{\mbox{$\frac12$},\mbox{$\frac12$}}$}
   \psfrag{b}{$\ket{\mbox{$\frac12$},-\mbox{$\frac12$}}$}
   \psfrag{c}{$\ket{\mbox{$\frac32$},\mbox{$\frac32$}}$}
   \psfrag{d}{$\ket{\mbox{$\frac32$},-\mbox{$\frac32$}}$}
   \psfrag{h}{$\ket{\uparrow} \leftrightarrow \ket{T_h^\downarrow}$}
   \psfrag{e}{(a)}
   \psfrag{f}{(b)}
   \psfrag{j}{(c)}
   \psfrag{y}{(d)}
   \psfrag{z}{(e)}
   \psfrag{g}{20 nm}
   \psfrag{r}{100 nm}
{\hspace{-2cm}\includegraphics[width=1\columnwidth]{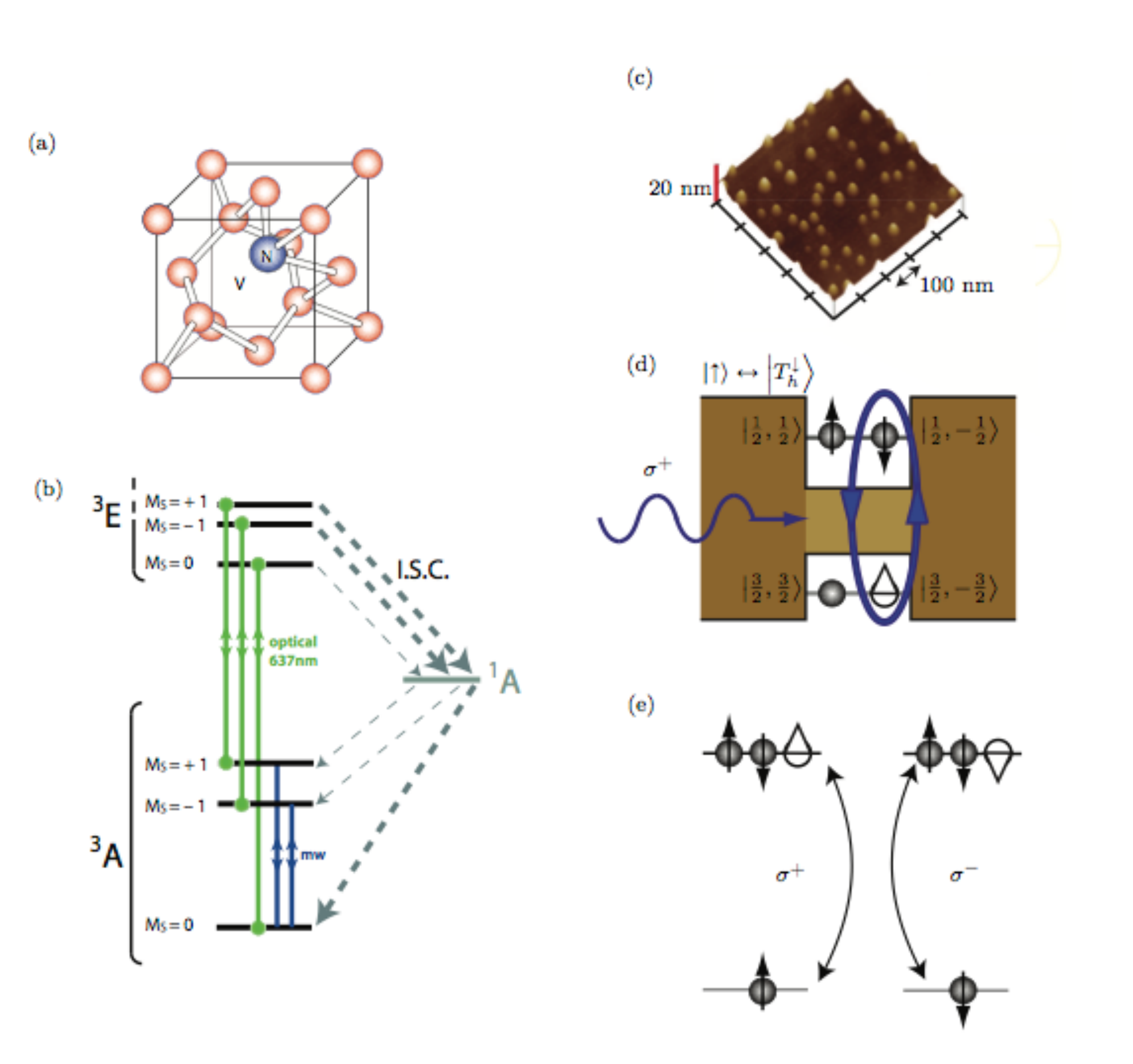}}}
  \end{psfrags}
\caption{Crystal structure (a) and electronic energy level diagram (b) of the NV$^-$ centre. (c) AFM micrograph showing the spontaneous formation of QDs when a smaller bandgap material (light brown) is deposited on a larger bandgap substrate (dark brown). (d) Confinement potential in a cut through one such QD, showing discrete bound states with angular momentum $\ket{J, J_z}$. Here the QD is doped with a single electron, and excitation with $\sigma^+$ light leads to  trion formation only for an initial $\ket{\uparrow}$ spin state. (e) Spin selection rules for both kinds of polarized light. [Figure partly adapted and reprinted with permission from \cite{kok10}.]}
\label{elecstruc}
\end{figure}

The second class are self-assembled nanostructures where confinement is naturally provided in all three dimensions. They are typically fabricated by molecular beam epitaxy: a large band-gap semiconductor forms the substrate and a smaller band-gap material is deposited on top (see Fig.~\ref{elecstruc}c). Under the right conditions, nanoscale islands form and subsequent overgrowth of the original material leads to three-dimensional confinement. The resultant discrete energy level structure of both conduction and valence bands is shown in Fig.~\ref{elecstruc}d and allows the physics of small numbers of electrons and holes to be investigated. The spin properties of both types of carrier are essentially determined by the corresponding bulk properties. In group IV or III-V semiconductors, valence states have $p$ like orbitals, and can have total spin $J=3/2$ or $1/2$, whereas conduction states have $s$ orbital symmetry and have have $J=1/2$~\cite{harrison00}. The confinement splits the six possible hole bands into three discrete doublets. The heavy holes ($J=3/2, J_z=3/2$ for growth direction $z$) generally have the largest mass of the valence states and under confinement form the highest lying doublet. 

Analogous to bulk semiconductors, optical QDs can either be intrinsic (i.e. have full valence and empty conduction states), or be doped to generate single electron or single hole ground states. Both are promising qubits and have been investigated experimentally~\cite{hanson08,ramsay08, gerardot08}, and theoretically~\cite{calarco03, gauger08} though most work has focussed on the electron spin~\cite{mikkelsen07, xu07}.  Inter-band (or, more correctly, inter-state) transitions typically lie in the optical or near infra-red region and can have significant transition dipole matrix elements~\cite{basu97, kok10}. These optically active transitions are the essential ingredient for an electron-photon interface as discussed in \Sec{sec:photons}.

\subsection{Impurities in solids}

The confinement achieved in quantum dots is naturally found in certain impurity or defect states within certain materials, some of which provide ideal hosts for an associated electron spin. Desirable properties of the host material include a low natural abundance of nuclear spins (such as C, Si, Ge and certain II-VI -based materials) and weak spin-orbit coupling.

In the semiconductor industry, silicon is doped with phosphorous in order to increase the bulk electron concentration, however at low temperatures the electron associated with the donor becomes bound to the P impurity. In his influential proposal~\cite{kane98}, Kane suggested using arrays of P donors in silicon (Si:P) as nuclear spin qubits, whose interactions could be tuned by electrical top-gates which change the wavefunction of the bound electron. This, and related proposals for Si:P quantum computing~\cite{skinner03, morton:clustersi}, are well supported by a number of findings and achievements made over the past decade: amongst others a) control of P-donor placement in silicon with atomic precision using scanning probe methods~\cite{ruess04}; b) manipulation of donor wavefunctions through electrical gates~\cite{calderon06 ,lansbergen08, bradbury06}; c) single spin detection in silicon~\cite{morello10}; and d) measurement of very long electron and nuclear coherence times, 0.6 and 3~seconds respectively, within a $^{28}$Si isotopically-enriched environment~\cite{morton:qmemory}; (A.M. Tyryshkin and S. A. Lyon, unpublished observation).

Other donors in silicon possess properties that make them of interest as electron spin qubits~\cite{stoneham03}. For example bismuth has a large nuclear spin ($I=9/2$) offering a large Hilbert space for storing quantum information~\cite{george10,morley10}, and its large hyperfine coupling ($A=1.475$~GHz) gives a zero-field splitting that may be useful for coupling to superconducting resonators~\cite{schuster10,kubo10}.

The paramagnetic nitrogen-vacancy NV$^-$ centre in diamond has an $S=1$ ground state, with a zero field splitting between the $m_s=0$ and $m_s=\pm1$ states of $\sim2.88$~GHz (see Figure~\ref{elecstruc} a and b). It exhibits coupling to surrounding \cthirteen~nuclei as well as the local nitrogen nuclear spin~\cite{jelezko04} and possesses a highly advantageous optical transition which enables initialisation and single-spin measurement~\cite{jelezko03} as discussed in detail in \Sec{sec:photons}. In addition, NV$^-$ centres offer the benefit of long coherence times at room temperature --- 1.8~ms in $^{12}$C-enriched diamond~\cite{balasubramanian09} --- which permits the measurement of coupling between NV$^-$ centres separated distances as long as 100~\AA~\cite{neumann10} (see also Box on Single-Spin Electron Spin Resonance).

There are other impurity spins with an associated optical transition which are at earlier stages of investigation, such as fluorine donors in II-VI semiconductors~\cite{sanaka09}, while rare-earth impurities in glasses are being investigated as optical memories for quantum repeaters~\cite{guillot08}.

\subsection{Molecular electron spin}
Molecules offer highly reproducible components which can host electron spin and can be coupled together using the techniques of organic and inorganic chemistry. Simple organic radicals, such as those based on nitroxide radicals, are used extensively in the field of spin-labelling for distance measurements in biological molecules~\cite{jeschke07}. Their electron spin coherence times are limited to  1--10~\mus~in typical environments, rich in nuclear spins from hydrogen nuclei, and can be extended to $\sim$~100 \mus~for dilute spins in deuterated environments at 40~K~\cite{lindgren97}. 

Fullerenes, such as \csixty~and \ceightytwo, act as a molecular-scale trap for atoms or ions, shielding the electron spin of the caged species from the environment. Such \emph{endohedral fullerenes}\footnote{The notation $M$@C$_{xx}$ is  used  to indicate the species $M$ is held within the C$_{xx}$ fullerene cage.}  based on group III ions such as Sc-, Y- and La@\ceightytwo~possess \ttwo~times in excess of 200~\mus~under optimised conditions~\cite{brown10}. In the case of the remarkable \natc~molecule, atomic nitrogen occupies a high-symmetry position at the centre of the cage, leading to an $S=3/2$ electron spin with the longest coherence times of any molecular electron spin: 80~\mus~at room temperature rising to 500~\mus~at temperatures below 100~K~\cite{mortoncs2, mortonprb}.

The organic radical created by X-ray irradiation of malonic acid crystals has been a standard in EPR/ENDOR spectrosocopy for many decades~\cite{mcconnell60}, and has also been used to explore methods of controlling coupled electron and nuclear spins qubits with a strong anisotropic hyperfine interaction~\cite{mehring03, mitrikas10}. Offering a large set of non-degenerate transitions, the high-spin ground state of many single-molecule magnets (SMMs) is capable in principle of hosting quantum algorithms such as Grover's search~\cite{leuenberger01}. Electron spin coherence (\ttwo) times up to a few microseconds have been measured~\cite{ardavan07}, permitting Rabi oscillations in the electron spin to be observed~\cite{mitrikas08}. Despite their relatively short coherence times, these tuneable systems may provide useful testbeds to explore quantum control of multi-level electron spin systems.

\subsection{Spins of free electrons}
Finally, it is possible to use free electrons as spin qubits. For example, using a piezeoelectric transducer over a semiconductor heterostructure, surface acoustic waves (SAWs) can be launched into a 2DEG, such that each SAW minimum contains a single electron~\cite{barnes00}. For more extreme isolation, electrons can be made to float above the surface of liquid helium, bound by their image charge. They can be directed around the surface by electrical gates beneath with very high efficiency~\cite{sabouret08}, for controlled interactions and measurement, meanwhile their spin is expected to couple very weakly to the varying electrical potentials~\cite{lyon06}.

\section{Electron spin - nuclear spin coupling}
\label{sec:nucspins}

\subsection{Electron spins as a resource for nuclear spin qubits}
\label{sec:efornucs}
Since the beginning of experimental studies into quantum information processing, nuclear spins and nuclear magnetic resonance (NMR) have provided a testbed for quantum control and logic~\cite{cory97, gershenfeld97}. NMR can still claim to have hosted the most complex quantum algorithm to date through the 7-qubit implementation of Shor's factoring algorithm~\cite{vandersypen01}. However, the weak thermal nuclear spin polarisation at experimentally accessible temperatures and magnetic fields has limited the scalability of this approach, which relies on manipulating the density matrix to create states which are pseudo-pure~\cite{cory97, gershenfeld97} and thus provably separable~\cite{braunstein99}. A notable exception (albeit of limited scalability) is the use of a chemical reaction on parahydrogen to generate two nuclear spin states with a purity of $\sim0.92$~\cite{anwar04}.

The magnetic moment of the electron spin is about two thousand times greater, bringing several key benefits to nuclear spin qubits: a) enhanced spin polarisation; b) faster gate manipulation time (ten nanoseconds for a typical electron spin single qubit rotation, rather than ten microseconds for the nuclear spin); and c) more sensitive detection, either via bulk magnetic resonance, or the more sensitive electrical or optical methods described in this review.

The general spin Hamiltonian for an electron spin $S$ coupled to one or more nuclear spins $I_i$ in a magnetic field $B$ is, in angular frequency units:
\beq
\mathcal{H}=\frac{g_e\mu_B}{\hbar}\vec{S}\cdot\vec{B}+\sum_i\left(\gamma_{n,i}\vec{I}_i \cdot\vec{B}+\vec{S}{\bf A}_i \vec{I}_i \right)
\eeq
where $g_e$ is the electron g-factor, $\mu_B$ the Bohr magneton, $\gamma_n$ the nuclear gyromagnetic ratio, and ${\bf A}_i$ the hyperfine coupling tensor between the electron and nuclear spin. Additional terms, such as a zero-field splitting term $\vec{S}{\bf D}\vec{S}$ may appear in higher spin systems, such as NV$^-$ centres in diamond.

For an electron and nuclear spin pair, this leads to four levels, separately addressable through resonant pulses and typically in the microwave regime (10--100~GHz) for the electron spin and in the radiofrequency regime (1--100~MHz) for the nuclear spin (see \Fig{fig:qmemory}B). By controlling the phase, power and duration of the pulse, qubit rotations about an axis perpendicular to the applied field are performed. Couplings that are stronger than the bandwidth of a typical pulse can be exploited to perform controlled-NOT (CNOT), or similar operations, through a selective microwave or rf pulse. Weaker couplings can be used to perform conditional logic through a combination of pulses and delays, exploiting the difference in the time evolution of one spin depending on the state of the other.

Electron spin polarisation can be indirectly (and incoherently) transferred to surrounding nuclear spins through a family of processes termed dynamic nuclear polarisation, reviewed extensively elsewhere~\cite{maly08,barnes08}. For strongly coupled nuclear spins, electron spin polarisation can be transferred directly through the use of selective microwave and rf pulses --- such a sequence forms the basis of the Davies electron nuclear double resonance (ENDOR) spectroscopic technique~\cite{davies74, mendor}. A complementary approach is to apply algorithmic cooling, which exploits an electron spin as a fast relaxing heat bath to pump entropy out of the nuclear spin system~\cite{ryan08}. The use of an optically excited electron spin (such as a triplet) can be advantageous as i) it offers potentially large spin polarisations at elevated temperatures and ii) the electron spin, a potential source of decoherence, is not permanently present~\cite{kagawa09}.

Given an isotropic electron-nuclear spin coupling of sufficient strength, it is possible to perform phase gates ($z$-rotations) on nuclear spin qubits on the timescale of an electron spin $2\pi$ microwave pulse, which is typically $\sim 50$~ns. The pulse must be selective on an electron spin transition in one particular nuclear spin manifold; hence, a weaker hyperfine coupling will necessitate a longer, more selective, microwave pulse. This kind of geometric Aharanov-Anandan phase gate~\cite{aharonov87} was experimentally demonstrated using \natc~and Si:P donor systems, and used to dynamically decouple nuclear spins from strong driving fields~\cite{morton:bangbang, tyryshkin06}. Given an anisotropic hyperfine coupling, the nuclear spin gate can be generalised to an arbitrary single-qubit rotation using a combination of microwave pulses and delays~\cite{mitrikas10, khaneja07}. For multiple coupled nuclear spins and a correspondingly large Hilbert space, more elaborate control is needed~\cite{hodges08}, for example using gradient ascent pulse engineering~\cite{khaneja05}. These methods exploit an effect termed electron spin echo envelope modulation (ESEEM) in the ESR community~\cite{hahn1952 ,dikanov92}.

The weak and `always-on' coupling between two nuclear spins is another limitation of a nuclear spin-only NMR approach. Nuclear spin interactions can be decoupled through refocussing techniques -- for example using the ultrafast nuclear spin manipulations described above -- however, methods for gating such interactions have also been explored. An example is the proposal of exploiting mutual coupling between two nuclear spins to an intermediate electron spin which is optically excited~\cite{schaffry10}. The mediator is diamagnetic in its ground state, such that the interaction between the two nuclei is effectively off. However, an optical pulse can excite a triplet state ($S=1$) in the mediator, which can be manipulated using microwave pulses to produce gates of maximal entangling power between the two coupled nuclear spins. Preliminary ENDOR experiments on candidate functionalised  fullerene molecules indicate that the key parameters of triplet polarisation, relaxation rate and hyperfine coupling, are in the correct regime to permit this kind of gate~\cite{schaffry10}.

Given the above, it is clear that quantum logic between an electron and nuclear spin can also be performed. Entangling operations between and electron and nuclear spin have been demonstrated in irradiated malonic acid crystals~\cite{mehring03}, and in the \natc~molecule~\cite{mehring04}; however in both cases the spins were in a highly mixed initial state and so only pseudo-entanglement was generated --- the states were fully separable. Nevertheless, Ref.~\cite{mehring03}  demonstrates an elegant way to perform density matrix tomography through the application of varying phase gates to the electron and nuclear spin, and a procedure which, if applied with high fidelity to spins at higher magnetic fields and lower temperatures, would lead to electron-nuclear spin entanglement.

\subsection{Nuclear spin quantum memory}

\begin{figure}[t]
{\includegraphics[width=\columnwidth]{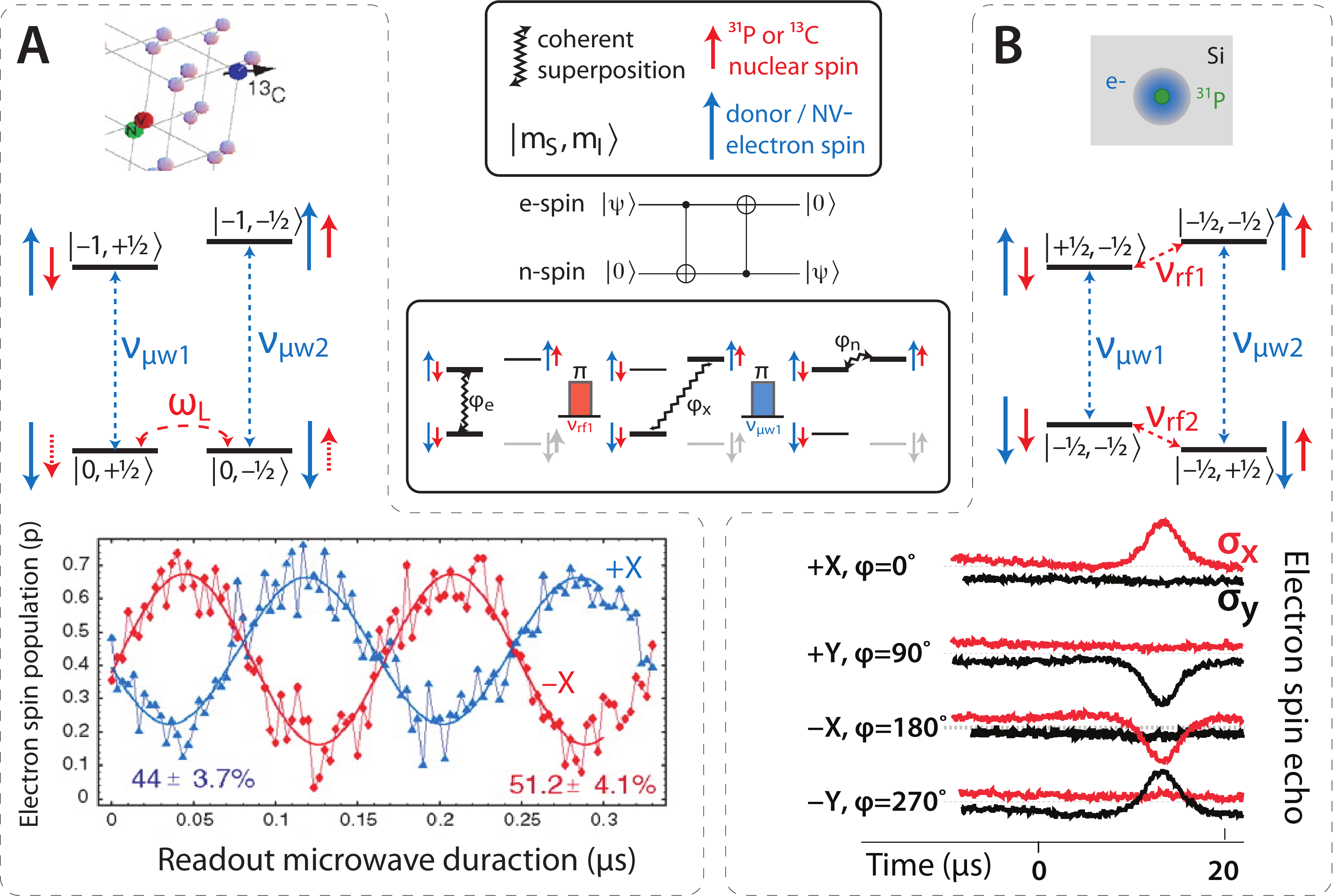}}
\caption{Two methods for coherently storing an electron spin state into a local, coupled nuclear spin. In both cases a SWAP operation is performed through two controlled-NOT (C-NOT) gates, which are achieved through a selective spin flip. (A) Using the NV$^-$ centre in diamond, it is possible to achieve the nuclear spin flip by exploiting the natural precession of the nuclear spin in the $m_S=0$ electron spin manifold. A weak ($\sim 20$~G) magnetic field is applied perpendicular to the quantisation axis which is defined by the electron spin and orientation of the defect, causing an evolution ($\omega/2\pi \sim 0.3$~MHz) in the nuclear spin between $\ket{\uparrow}$ and $\ket{\downarrow}$. (B) The alternative, shown using P-donors in \sitwoeight, is to drive a nuclear spin flip directly with a resonant radiofrequency pulse. In each case, electron spin coherence is generated, stored in the nuclear spin, and then retrieved some time later which is considerable longer than the electron spin \ttwo. Observation of the recovered electron spin coherence can be achieved through Rabi oscillations, or by directly observing a spin echo, depending on the nature of spin measurement used. (Panel A adapted from Reference~\cite{dutt07} with permission from AAAS; Panel B adapted from Reference~\cite{morton:qmemory}, with permission from Macmillan Publishers Ltd.)
}
\label{fig:qmemory}
\end{figure}

We may reverse the question, and ask how coupling to nuclear spins may offer advantages to electron spin qubits. One key advantage of the weak magnetic moment of nuclear spins is their correspondingly longer relaxation times \tone~(typically seconds to hours)~and \ttwo~(typically seconds), motivating the use of nuclear spins as quantum memories to coherently store the states of electron spin qubits. This has been achieved using a) NV$^-$ centres and neighbouring \cthirteen~nuclei in diamond, exploiting the near-degeneracy of the nuclear spin levels in the $m_S=0$ manifold~\cite{dutt07}; and b) P-donor nuclear spins in isotopically purified \sitwoeight, where the nuclear spin is directly excited using a radiofrequency pulse~\cite{morton:qmemory}. Both experiments are summarised in \Fig{fig:qmemory}. 

The lifetime of the quantum memory is determined by the nuclear spin decoherence time \ttwon, which was found to be $\gg20$~ms in the case of \cthirteen~in the NV$^-$ diamond system at room temperature~\cite{dutt07}, and over 2 seconds for the \ptone~donor nuclear spin. This approach has since been applied to other electron-nuclear spin systems such as the \natc~molecule (\ttwon$ = 140$~ms at 10~K) (R. M. Brown, A. M. Tyryshkin, K. Poerfyrakis, E. M. Gauger, B. W. Lovett, et al., unpublished observeration) and the substitutional nitrogen P1 centre in diamond (\ttwon$ = 4$~ms at room temperature). 

The large $I=9/2$ nuclear spin of the \biiso~donor in silicon offers a large Hilbert space to store multiple electron spin coherences. Despite the wide range of nuclear transition frequencies (between 200 and 1300~MHz at X-band), it is possible to implement the same coherence transfer sequence applied previously to P-donors~\cite{george10}. A further strength of this system is the ability to optically hyperpolarise the \biiso~nuclear spin~\cite{sekiguchi10,morley10}, as was previously shown for the P-donor~\cite{mccamey09, yang09}.

An important limitation of the use of this kind of nuclear spin quantum memory is that the nuclear spin coherence time is bounded by the electron spin relaxation time: \ttwon~$\leq$~2\tonee~\cite{morton:qmemory}. In many systems \tonee~can be made very long, for example by operating at low temperatures, however the ability to remove the electron spin, for example through optical or electrical means, would simply remove this limit~\cite{schaffry10, morton:clustersi}.

\section{Electron spin - optical photon coupling}
\label{sec:photons}

\subsection{Mechanisms and candidate systems}

There are various methods by which spin can couple to an optical transition in certain materials, exploiting some kind of spin-selective interaction with light. This interaction enables the initialisation, manipulation and measurement of single spins using optical techniques, which we shall review in this section.

NV$^-$ centres (see \Fig{elecstruc}a,b) possess an optically active level structure that has a number of fortuitous properties. In particular, there is an intersystem crossing (ISC) which can take place between the excited $^3$E state and a metastable singlet $^1$A state, and the rate of ISC is three orders of magnitude faster for the $m_s=\pm1$ excited states than for $m_s=0$. Crucially, optical cycles between the ground and excited triplets state are essentially spin conserving, and relaxation from the $^1$A state back to the triplet ground state  $^3$A occurs with greatest probability to the $m_s=0$ state~\cite{manson06}.

Spin selectivity in QDs is illustrated in \Fig{elecstruc}d,e. A left (right) circularly polarized light pulse propagating in the $z$ direction carries an angular momentum of $+\hbar$ $(-\hbar)$ along the $z$ axis, and if it is resonant with the heavy-hole to electron transition will only excite the transition that has a net angular momentum loss (gain) of $\hbar$. If the QD is doped with a single electron, then this leads to a spin dependent optical transition for a given circular polarization. For example, if $\sigma_+$ light is incident on the sample propagating in the $z$ direction, then a state consisting of two electrons and one hole (known as a trion) is only created from the $J_z = +1/2$ level. This kind of `Pauli blocking' forms the basis of methods for optical initialization, readout and manipulation of spins in QDs.

Optical coupling to spin qubits has been explored in other systems, such as donors in silicon~\cite{yang09}, but we will focus here on NV$^-$ and QD systems to illustrate the techniques and opportunities for coupling electron spins and photons in the solid state.



\subsection{Electron spin initialisation}

Electron spins can become highly polarized in strong magnetic fields and low temperatures (e.g. 90\% polarization at 2~K and 4~T). However, it is possible to use optical colling to achieve similar polarizations at much lower magnetic fields and more accessible temperatures.
The initialisation of NV$^-$ centre spins at room temperature is possible because cycling the 637~nm optical transition largely preserves the electron spin state. As described above, the $m_s=0$ state has a very low probability of undergoing ISC when in the excited state $^3$E, while the $m_s=\pm1$ states have some change of crossing to $^1$A, which will relax to $m_s=0$. A few cycles is enough to generate a large spin polarisation ($\sim90\%$) in the $m_s=0$ state~\cite{wrachtrup06,neumann10}. Unfortunately, no method to increase this polarisation closer to 100$\%$ has yet been identified; the difficulty is that the optical transitions are not perfectly spin-conserving and so there is a finite change of a spin-flip on each optical cycle~\cite{manson06}.

Atat\"ure {\it et al.}~\cite{atature06} demonstrated laser induced spin polarisation in a InAs/GaAs QD~\cite{xu07}. They use a $\sigma^-$ laser to depopulate the $|\downarrow\rangle$ spin level, promoting population to the trion above, see Fig.~\ref{elecstruc}e. The trion decays primarily back to the original state, but there is a small probability that it goes to the other low lying spin level ($|\uparrow\rangle$) via a spin flip Raman process that arises due to light-heavy hole mixing. However, any population in $|\uparrow\rangle$ will remain, since the pump has no effect on it. In this way, polarization builds up as eventually the $|\downarrow\rangle$ is completely emptied, so long as there is no direct spin flip mechanism with a rate comparable with the forbidden decay rate. In zero magnetic field, interaction with the nuclear spin ensemble does lead to spin flips and only in an applied magnetic field can such flips be sufficiently suppressed. The measurement of polarization is made by observing the change in transmission of the probe laser: once the spin is polarized no more absorption can occur. 
However, a sufficiently sensitive measurement can only be obtained by exploiting a differential transmission technique~\cite{alen03, hogele05}.


Other methods of spin initialization rely on the generation of a polarized exciton in diode-like structure that permits the preferential tunneling of either an electron of hole. Both electrons~\cite{kroutvar04} and holes~\cite{ramsay08} can be prepared in this way.



\subsection{Electron spin measurement}

The spin-selective ISC in the excited $^3$E state enables the measurement of the spin state of a single NV$^-$ centre. The lifetime of the dark $^1$A state ($\sim250$~ns) is over an order of magnitude longer than that of $^3$E, such that the fluorescence intensity of the centre is reduced when ISC can occur~\cite{manson06}. The act of measurement (cycling the $^3$A-$^3$E optical transition to observe fluorescence) itself serves to re-initialise the spin in the $m_s=0$ state, so the signature of the spin measurement is a  $\sim20\%$ difference in fluorescence intensity in the first 0.5~\mus~or so of optical excitation. This means that each measurement must be repeated many times in order to build up good contrast between the different spin states. Experiments showing single spin measurement are therefore a time-ensemble average --- in contrast to the spin-ensemble average typical of ESR experiments --- and refocusing techniques must be employed required to remove any inhomogeneity~\cite{jelezko03}.

Methods to improve the efficiency of the measurement are being actively explored, for example using \cthirteen~nuclear spins~\cite{jiang09} or the \nfourteen~nuclear spin~\cite{steiner10} as ancillae for repeated measurement. The electron spin state is copied to ancilla(e) nuclear spin(s), and then measured\footnote{This is not a cloning of the electron spin state (which is forbidden), but rather a C-NOT operation such that $\alpha\ket{0_e0_n}+\beta\ket{1_e0_n}\rightarrow \alpha\ket{0_e0_n}+\beta\ket{1_e1_n}$}. After some time ($<1$~\mus ), any useful information on the electron spin state ceases to be present in the fluorescence and the electron spin is back in the $m_s=0$ state; the ancilla state can be mapped back to the electron spin and the measurement repeated. Crucially, the coherent state of the coupled nuclear spin is not affected by the optical cycling that forms part of the electron spin measurement~\cite{dutt07, jiang08}.

A range of techniques for single spin measurement have been explored for QDs. Several of these rely on the modification of refractive index that occurs close to (but not precisely at) optical resonance. By detuning a probe laser from resonance, absorption is suppressed and this dispersive regime can be exploited. Imagine, for example, a single electron spin polarized in the $\ket{\uparrow}$ state. It will only interact with $\sigma^+$ light, and on passing through the quantum dot region, light of this polarization will experience a slightly different optical path length to an unaffected $\sigma^-$ beam. In order to enhance the difference in path lengths experienced, an optical microcavity can be exploited so that on average the light beam passes many times through the dot region. Linearly polarized light is composed of equal amounts of $\sigma^+$ and $\sigma^-$, and will therefore {\it rotate} on passing through the sample. The degree of rotation will be directly dependent on the magnitude and direction of the confined spin. The sensitivity of a typical measurement is improved by recording the degree of rotation as a function of detuning from resonance, and fitting the resulting curve with the initial spin polarization as a variable fitting parameter.

\begin{figure}[t]
\scriptsize
  \begin{psfrags}
   \psfrag{u}{$\sigma^-$}
   \psfrag{v}{$\sigma^+$}
   \psfrag{G}{Gate voltage (mV)}
   \psfrag{D}{Dispersive signal (a.u.)}
      {\includegraphics[width=0.75\columnwidth]{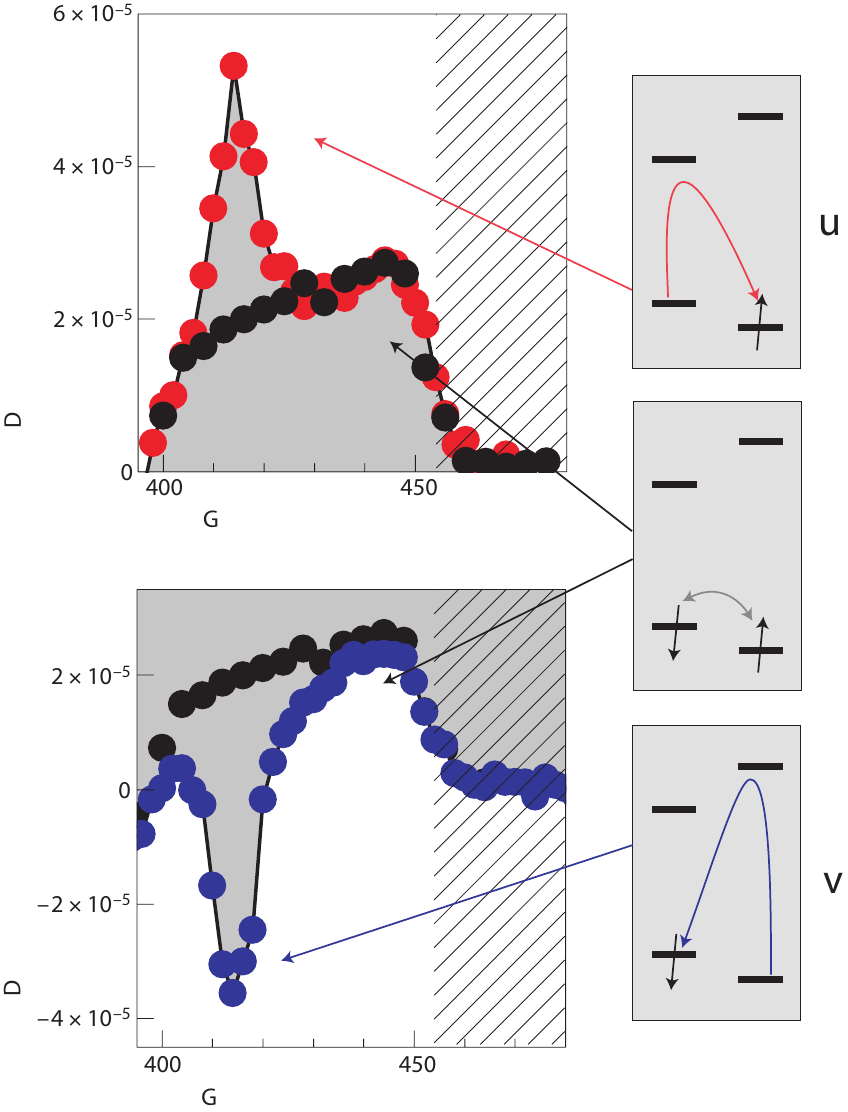}}
        \end{psfrags}
\caption{Experimental results of a preparation-measurement sequence for a spin in a QD. A preparation pulse is first applied with circular polarization which results in spin pumping on resonance. Both figures shows a dispersive Faraday rotation (FR)measurements as a function of an applied gate voltage that is used to change the detuning of the preparation laser from resonance through the Stark shift it generates. The upper (lower) figure show results using a $\sigma^-$ ($\sigma^+$) preparation pulse. The black dots in each case represent the signal for a far detuned preparation laser, with no spin cooling. The red (blue) dots show a preparation laser that is at resonance for a gate voltage of 415 mV, for the transitions shown on the right. A peak or dip in the FR signal is seen at resonance, which is a measurement of the prepared spin state. [Reprinted  from Ref.~\cite{atature07}, with permission from Macmillan Publishers Ltd.]}
\label{atature1fig}
\end{figure}

Berezovsky {\it et al.}~\cite{berezovsky06} demonstrated this spin-dependent polarization rotation using a reflection geometry (in this case the effect is called {\it magneto-optical Kerr effect}) to non-destructively read out a single spin following initialization in the $|\uparrow\rangle$ or $|\downarrow\rangle$ state. 
Atat\"ure {\it et al.}~\cite{atature07} performed similar measurements (see \Fig{atature1fig}), but using differential transmission to detect the rotation of polarization, which in this configuration is called Faraday rotation. 

State readout can also be achieved using a photocurrent technique, in which a trion that has been spin selectively created is allowed to tunnel from a QD in a diode structure~\cite{kroutvar04, ramsay08}.

\subsection{Electron spin manipulation}

Once initialization and measurement are established, the stage is set for the observation of controlled single spin dynamics. 
In the case of electron spins which are sufficiently long-lived, this can be performed by coupling the spin to a resonant microwave field, such as that used to observed coherent oscillations in a single NV$^-$ electron spin between the $m_s=0$ and the $m_s=\pm1$ levels~\cite{jelezko03}. 
Alternatively, by applying a magnetic field perpendicular to the direction of the initialization and measurement basis, it is possible to observe single spin precession as a function of time between the initialization `pump' and Kerr rotation `probe'~\cite{mikkelsen07}. For controlled rotation around a second axis, an optical `tipping' pulse can be applied~\cite{berezovsky08}. This pulse is circularly polarized, and cycles round one of the spin selective transitions shown in Fig.~\ref{elecstruc}e. If the pulse is resonant, and completes a cycle from spin to trion and back, a relative $\pi$ phase shift is accumulated between the two spin states\footnote{This is analogous to the nuclear spin phase gate described in \Sec{sec:efornucs} performed by driving the electron spin around a complete cycle. Similar phase gates can also be achieved in a photodiode system~\cite{ramsay08}.}. This corresponds to a rotation around the measurement basis axis, perpendicular to the applied field. Controlled rotation around two axes is sufficient for arbitrary single qubit operations~\cite{nielsen00}, though these experiments do not demonstrate how to stop the magnetic field precession. Greilich {\it et al.}~\cite{greilich09} take the first step to overcoming this problem, first  by demonstrating that an arbitrary angle phase gate can be achieved by detuning the cycling laser, and then timing two $\pi/2$ phase gates such that a controlled field precession occurs in between them.
Experimental progress on more than one dot has naturally been slower, but theoretical work has shown that optical manipulation of strongly interacting quantum dot molecules can lead to spin entanglement~\cite{calarco03} with relatively simple pulses~\cite{nazir04}. 

A potential drawback of using trion states to manipulate single spins is that the charge configuration of the system can change significantly during a control pulse. This can lead to a strong phonon coupling~\cite{ramsay10} and to decoherence. However, by using chirped pulses or slow amplitude modulation, this decoherence channel can be strongly suppressed~\cite{lovett05, gauger08, gauger08b, roszak05}.

\subsection{Single shot measurement}

The measurements so far discussed have relied on weak dispersive effects in the case of QDs, or small changes to the fluorescence over a limited time window in the case of NV$^-$ centres. A full measurement of a single spin is a rather slow process 
and each of the experiments we have described prepare and measure the spin many times over to build up enough statistics to prove the correlation between the initialized state and the measured state. For most quantum computing applications, this will not be enough. Rather, high-fidelity single-shot determination of an unknown spin is required, and for this a much stronger spin-photon coupling must be engineered and exploited. 

The first step towards this ideal in QDs was a demonstration that spins can be measured at resonance. Under this condition, the effect of a QD on a photon is greater than it can be in a detuned, dispersive set-up. If a two level system is driven resonantly, and the coupling between photon and exciton is stronger than the exciton decoherence rate, then the eigenstates of the system become polaritons (states with both photon and exciton character). Moreover, the emission spectrum changes from a single Lorentzian to a `Mollow' triplet~\cite{mollow69}, with two side bands split from the central peak by the QD-photon coupling.\footnote{The triplet arises from a splitting of both upper and lower levels into doublets; a triplet results since two of the transitions are degenerate. If the transition to a third level from one of the two doublets is probed, two distinct lines are indeed observed~\cite{xu07a, boyle09}.}  It is a significant experimental challenge to observe the Mollow triplet, since the sidebands can be obscured by Rayleigh scattering of the incident light beam. However, Flagg {\it et al.}~\cite{flagg09} showed that it is possible, by sandwiching the QD between two distributed Bragg reflector mirrors. Laser light is coupled into the layer between the mirrors using an optical fibre, and confined there by total internal reflection. The emission is then observed in a perpendicular direction to reduce any interference from the excitation beam. This work was soon followed by an observation of a `quintuplet' by another group~\cite{vamivakas09} where the sidebands are spin-split in an applied magnetic field. However, this kind of measurement is not immune from spin flip Raman processes, which still occur at a faster rate than the readout can be performed. In a recent demonstration of single-shot optical measurement~\cite{atature10}, two QDs are used so that the initial spin and readout exciton are spatially located in different dots. The readout process then proceeds through a trion consisting of two electrons with parallel spins and a heavy hole, and from this state electron spin flips through Raman processes are not possible. 

In NV$^-$ centres, the emphasis has been on increasing the efficiency of optical coupling, for example using microcavities based on silicon microspheres~\cite{larsson09} or on-chip designs using GaP~\cite{barclay09}. Other approaches include the use of nanostructured diamond itself for optical confinement, for example diamond nanowires~\cite{babinec10}, or using photonic crystal cavities which can be detereministically positioned to maximise coupling with the NV$^-$ centre~\cite{englund10}.

\subsection{Single-photon coupling and entanglement}

\begin{figure}[t]
\footnotesize
  \begin{psfrags}
   \psfrag{a}{(a)}
   \psfrag{b}{(b)}
   \psfrag{p}{$D_1$}
   \psfrag{q}{$D_2$}
   \psfrag{0}{$\ket{\downarrow}$}
   \psfrag{1}{$\ket{\uparrow}$}
   \psfrag{f}{$\ket{T_h^\uparrow}$}
   \psfrag{e}{$\ket{T_h^\downarrow}$} 
     \psfrag{x}{$\omega$}
      {\includegraphics[width=\columnwidth]{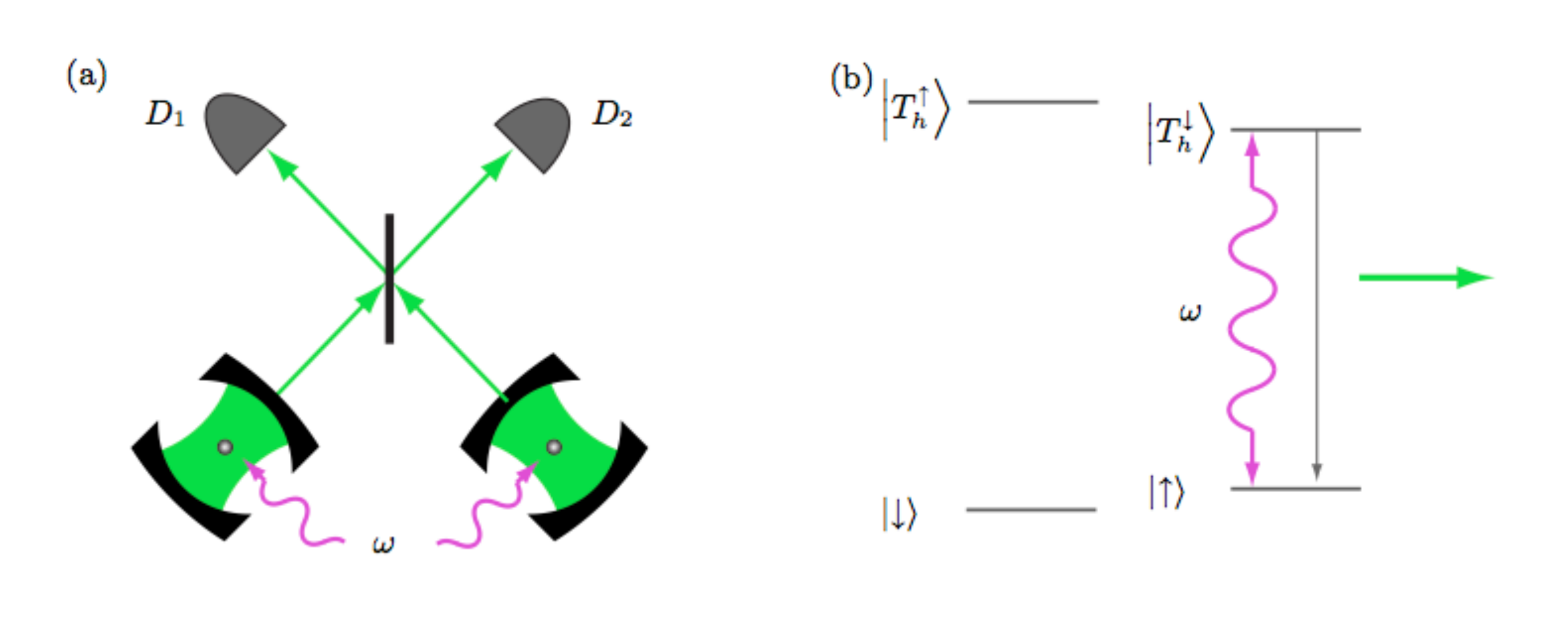}}
        \end{psfrags}
\caption{(a) Schematic diagram showing the experimental setup needed for the creation of entanglement by measurement. Two QDs are embedded in cavities to enhance emission into the modes shown. Two detectors $D_1$ and $D_2$ are placed downstream of a beamsplitter; detection of one photon heralds an entangled spin state. (b) Required level structure of the QD. [Figure adapted and reprinted with permission from \cite{kok10}.]}
\label{mbqc}
\end{figure}

A more ambitious goal is to achieve single shot measurement by detecting a single photon. If this could be achieved, it would permit the creation of spin entanglement between two matter-based electron spin placed at distant locations~\cite{benjamin09}. We will describe how this could be done using QDs (see Fig.~\ref{mbqc}), however similar schemes have proposed for NV$^-$ centres~\cite{barrett05a}. Imagine two QDs in different places with the kind of spin selectivity discussed above, and that polarized light is used to drive just one of the two transitions in both QDs. Each QD is first initialised in the state $(\ket{\uparrow}+\ket{\downarrow})/\sqrt{2}$ so that following optical excitation we have the state
\begin{equation}
\ket{\Psi} = \frac{1}{2}\left(\ket{\downarrow \downarrow} + \ket{\downarrow T_h^\uparrow} + \ket{T_h^\uparrow \downarrow} + \ket{T_h^\uparrow T_h^\uparrow}\right).
\end{equation}
This will eventually decay back to the ground state, with the emission of either no photons (corresponding to the first term on the RHS), one photon (second and third term), or two photons (last term). The QDs  are placed inside cavities such that the emission is almost always in the directions shown. If the detectors can discriminate between different photon numbers, but they register just one photon between them, then the system is projected into the state corresponding to the decay products of the second and third terms. Importantly, we cannot tell which one since each passes through a 50:50 beamsplitter before being detected and if the QDs are identical this erases any information about which path the photon takes before the beamsplitter. We are therefore projected into an entangled state of the form $(\ket{\uparrow\downarrow} + \exp(i\phi)\ket{\downarrow\uparrow})/\sqrt{2}$, with the phase factor determined by which detector fires. Several theoretical ideas for entanglement creation along these lines have been put forward over the last few years~\cite{cabrillo99, bose99, barrett05a, lim05, kolli06} and aspects have been demonstrated in ion traps and atomic ensembles~\cite{moehring07, laurat07}. The most important feature of this kind of entanglement creation is that is it heralded: we know when our operation works and when it fails simply by counting detection photons. 

By performing successful entangling measurements on adjacent pairs of spins, a large entangled resource can be built up. NV$^-$ centres possess a local, coupled nuclear spin which is immune to the optical excitation performed while attempting to create electron spin entanglement --- this allows a broker-client approach to efficiently building up larger entangled states~\cite{benjamin06}. Any quantum algorithm can be performed using such a resource simply by making single qubit measurements~\cite{raussendorf01, kok10}. 


Controlled entanglement of a single spin and a single photon would represent the ultimate interface of the two degrees of freedom\cite{yao05}. 
Such a static to flying qubit interconversion would allow for secure quantum communication~\cite{divincenzo00} and the construction of truly quantum networks~\cite{cirac97}. 
Certain key aspects such as strong dot-cavity coupling~\cite{yoshie04, reithmaier04,hennessy07} have been experimentally demonstrated, and recently (non-deterministic) entanglement between a single spin and a photon was detected in the NV$^-$ system~\cite{togan10}.






\section{Electron spin - charge coupling}
\label{sec:charge}


\subsection{EDMR}
The relationship between electrical conductivity and the spin states of conduction and localised electrons within a material has been exploited for some time in the technique of electrically detected magnetic resonance (EDMR)~\cite{schmidt66,lepine72, brandt04}. Although the effect is often very weak --- only a small fraction of a percent change in conductivity --- it offers an opportunity to detect and measure small numbers of electron spins ($<100$) by measuring the conductivity of nanostructured devices~\cite{mccamey06}.

Three primary mechanisms for EDMR have been used to measure electron spins in semiconductors: spin-dependent scattering~\cite{ghosh92, lo07}, spin-dependent recombination~\cite{mccamey06} and spin-dependent tunnelling~\cite{ryan09}.  Spin-dependent scattering is observed in MOSFET structures where the scattering cross-section of a 2DEG electron and a bound donor electron depends on their relative spin orientation (i.e.\ single vs.\ triplet)~\cite{ghosh92}. 

Spin-dependent recombination involves first optically generating charge carriers in the material, which then recombine via charge traps. Commonly, this technique involves \emph{two} species of charge trap, such as a dangling bond P$_{\rm b0}$ centre in silicon coupled, for example, to a P-donor~\cite{stegner06}. If the two trapped charges are in the singlet (vs the triplet) state, the P-donor will transfer its electron to the P$_{\rm b0}$ centre, and subsequently capture an electron from the conduction band. Recombination then takes place between the P$_{\rm b0}^-$ state and a hole, such that the overall process reduces the carrier concentration and thus the conductivity of the device.  An alternative is to use just one charge trap, such as the P-donor itself which ionises to the $D^-$ state when trapping a conduction electron~\cite{morley08}. Trapping is only possible when the conduction spin and trap spin are in a singlet state, which leads to a measurable change in the recombination rate if both spins are highly polarised, requiring high magnetic fields and low temperatures (e.g.\ $>3$~T and $<4$~K). This has the advantage of being able to measure donor spins which are not necessarily coupled to interface defects which lead to shorter relaxation times~\cite{paik10}.  


Various approaches have been proposed and explored to extend these ideas to the single-spin level --- in most cases this involves scaling the devices used down to a sufficiently small scale so that the behaviour of a single donor has a non-negligible impact on the conductivity of the device~\cite{beveren08, lo09}. It is also possible to use the donor nuclear spin as an ancilla for repetitive measurement to enhance the sensitivity~\cite{sarovar08}.

Although these methods could be used to ultimately measure a single spin state, they do so using a large number of electronic charges --- analogous to the way spins 
can be measured through an optical fluorescence or polarisation change detected using a large number of photons. We now examine how the state of a single charge can be coupled to an electron spin, offering a powerful electrical method for initializing, measuring and manipulating single spins. 

\subsection{Single charge experiments}



As discussed in Section~\ref{subsec:qds}, the charge state of lithographically defined quantum dots can be controlled by applying voltages to gate electrodes, and the isolation of just a single electron in a QD was achieved many years ago (see Ref.~\cite{kouwenhoven01} for a detailed review of achievements in QD electron control that happened in the 1990s). In the last decade, experiments have shown the conversion of information carried by single spins into measurable single charge effects, so-called \emph{spin-to-charge conversion}. Several excellent reviews of this topic already exist~\cite{hanson07, hanson08}, and so here we will touch on a few key early results, and then focus on some more recent achievements.

An early measurement of single spin dynamics that used a form of spin-to-charge conversion was achieved by Fujisawa {\it et al.} in 2002~\cite{fujisawa02}. They demonstrated that a single QD could be filled with one or two electrons by applying a particular voltage to the QD gate electrode, states which they termed artificial hydrogen and helium atoms. In the two electron case, the authors demonstrated that the number of charges escaping from the QD can be made proportional to triplet population, and in this way triplet to singlet relaxation times were probed. 

Single shot readout of a single spin was achieved two years later~\cite{elzerman04}, using a single gate-defined QD in a magnetic field that is tunnel coupled to a reservoir at one side, and has a quantum point contact (QPC) at the other side. A QPC is a one-dimensional channel for charge transport, whose conductance rises in discrete steps as a function of local voltage~\cite{wharam88, wees88}. 
By tuning the QPC to the edge of one of these steps, a conductance measurement can act as a sensitive charge detector. The QD is first loaded with a single electron by raising the potential of the plunger gate, which transfers a random spin from the reservoir. After a controlled waiting time, the voltage is lowered to a level such that the two Zeeman split spin levels straddle the Fermi surface of the reservoir. It is therefore only possible for the spin-up state to tunnel out, an event that can be picked up by the QPC. Such spin-dependent tunnelling lies at the heart of almost all spin-to-charge conversion measurements in these kinds of systems. By repeating the procedure many times, and for different waiting times, it is possible to measure the electron spin relaxation ($T_1$) time.

Coherence times ($T_2$) of electrons can be probed using a similar single shot read-out technique; Petta et al.~\cite{petta05} did this using a double QD loaded with two electrons. Depending on the relative potentials of the two dots, the electrons can either be on the same dot (labelled (1,1)) or on different dots (2, 0). A nearby QPC is sensitive to charge on one of the two dots and so can distinguish these two situations. Importantly, for (2, 0) the electrons must be in the singlet state, due to Pauli's exclusion principle. This provides means of initializing, and reading out (1, 1) states: a triplet (1,1) will not tunnel to (2, 0) when the dot detuning is changed to make (2, 0) the ground state, but a singlet will. In this way, the dynamics of a qubit defined by the singlet and the triplet state with zero spin projection can probed. Since these two states differ only by a phase this leads to a measurement of the dephasing $T_2$ time, which is found to be primarily dependent on the interactions with the nuclear spins on the Ga and As atoms. Though it is possible to remove the effects of static nuclear spins through rephasing techniques, nuclear spin fluctuations limit $T_2$ to the microsecond regime.

More recent achievements have focussed on the manipulation of electron spins using ESR, permitting the observation of Rabi oscillations. For example, in~\cite{koppens06}, an oscillating magnetic field is applied to a two QD system by applying a radio-frequency (RF) signal to an on-chip coplanar stripline. The resulting Rabi oscillations can be picked up by measuring the current through the device following the RF pulse; Pauli's exclusion principle means that the current is maximized for antiparallel spins. In an alternative approach, a magnetic field gradient was applied across a two QD device using a ferromagnetic strip so that field strength depends on  the equilibrium position of the electron charge~\cite{pioroladriere08}. The application of an oscillating electric field induces an oscillation of the electron charge --- the electron spin then `feels' a periodic modulation of the magnetic field. If the modulation is at a frequency corresponding to the spin resonance condition, then Rabi oscillations occur.

The last two or three years have seen a remarkable improvement in our understanding and control of the primary electron spin decoherence mechanism: interactions with the GaAs nuclear spin bath. For example, the nuclear spin bath can be polarized by transferring angular momentum from the electron spins~\cite{reilly08}, resulting in a narrower field distribution. The nuclear spin field can also adjust itself so that the condition for electron spin resonance is satisfied for a range of applied fields~\cite{vink09}; this may also result in a reduction in electron spin dephasing by narrowing the nuclear field distribution. Remarkably, fast single shot readout of an electron spin has permitted the static nuclear field to be measured and tracked through the real time impact is has on the singlet-triplet qubit discussed above~\cite{barthel09}.

\begin{figure}[t]
{\includegraphics[width=\columnwidth]{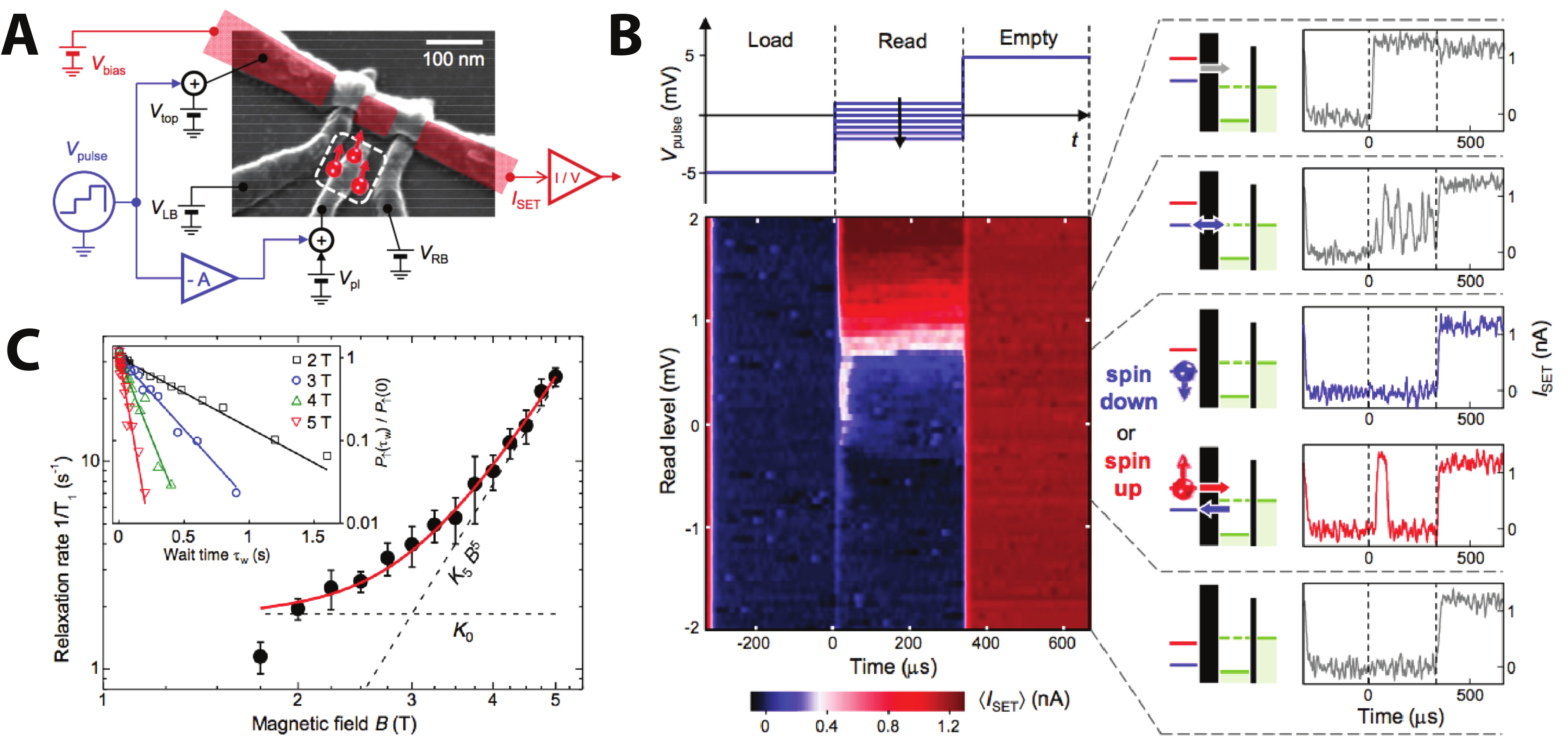}}
\caption{Single-shot single spin to charge conversion in silicon using a MOS single electron transistor (SET). A) P-donors are implanted within the vicinity of the SET island, underneath a plunger gate $V_{pl}$ which controls their potential with respect to the SET. Statistically, one expects $\sim3$ donors to be sufficiently tunnel-coupled to the SET. B) An electron can be loaded onto the donor by making the chemical potential $\mu_{\rm donor}$ much lower than that of the SET island $\mu_{\rm SET}$. Placing  $\mu_{\rm SET}$ in between the spin-up and spin-down levels of the donor enables spin-dependent tunnelling of the donor electron back onto the SET island. C)  The electron spin relaxation time \tone~is measured by loading an electron (with a random spin state) onto the donor, followed by a measurement of the spin some variable time later. The $B^5$ magnetic field dependence, and absolute values of the \tone~measured are consistent with a P-donor electron spin (S. Simmons, R. M. Brown, H. Riemann, N. V. Abrosimov, P. Becker, et al., unpublished observation). (Reprinted with permission from Reference~\cite{morello10} with permission from Macmillan Publishers Ltd.)}
\label{morello}
\end{figure}

Removing nuclear spins altogether is also possible, for example using QDs in silicon- or carbon-based devices such as those fabricated using graphene or carbon nanotubes~\cite{biercuk05}. 
Demonstration of a single spin to charge conversion has now been observed in silicon~\cite{morello10} (as described in \Fig{morello}), while the demonstration of the coherent exchange of electrons between two  donors~\cite{verduijn10} may provide a route to generating spin entanglement.


\section{Electron spin and superconducting qubits}

Qubits based on charge-, flux-, or phase-states in superconducting circuits can be readily fabricated using standard lithographic techniques and afford an impressive degree of quantum control~\cite{clarke08}. Multiple superconducting qubits can be coupled together through their mutual interaction with on-chip microwave stripline resonators~\cite{niskanen07}. These methods have been used to demonstrate the violation of Bell's inequalities~\cite{ansmann09} and to create three-qubit entangled states~\cite{neeley10,dicarlo10}. 

The weakness of superconducting qubits remains their short decoherence times, typically limited to a few microseconds depending on the species of qubit~\cite{clarke08}, for example, coherence times in the three-qubit device of Ref~\cite{dicarlo10} were limited to less than a microsecond. Electron spins, in contrast, can have coherence times up to 0.6~s in the solid state, and typically operate at similar microwave frequencies as superconducting qubits. The possibility of transferring quantum information between superconducting qubits and electron spins is therefore attractive.

Although it is possible to convert a superconducting qubit state into cavity microwave photon~\cite{wallraff04}, achieving a useful strong coupling between such a photon and a single electron spin appears beyond capabilities of current technology\footnote{A superconducting cavity could be used for single spin ESR (see Box: Single-spin electron spin resonance), however, this application places fewer demands on the coupling and cavity/spin linewidths than a quantum memory}. Instead, the spin-cavity coupling (which scales as $\sqrt{N}$ for $N$ spins) can be enhanced by placing a larger ensemble of spins within the mode volume of the cavity, thus ensuring that a microwave photon in the cavity is absorbed into a collective excited state of the ensemble. The apparent wastefulness of resources (one photon stored in many spins) can be overcome by using holographic techniques to store multiple excitations in orthogonal states within the ensemble~\cite{wesenberg09}, as has been explored extensively in optical quantum memories~\cite{reim10}. 

\begin{figure}[t]
{\includegraphics[width=\columnwidth]{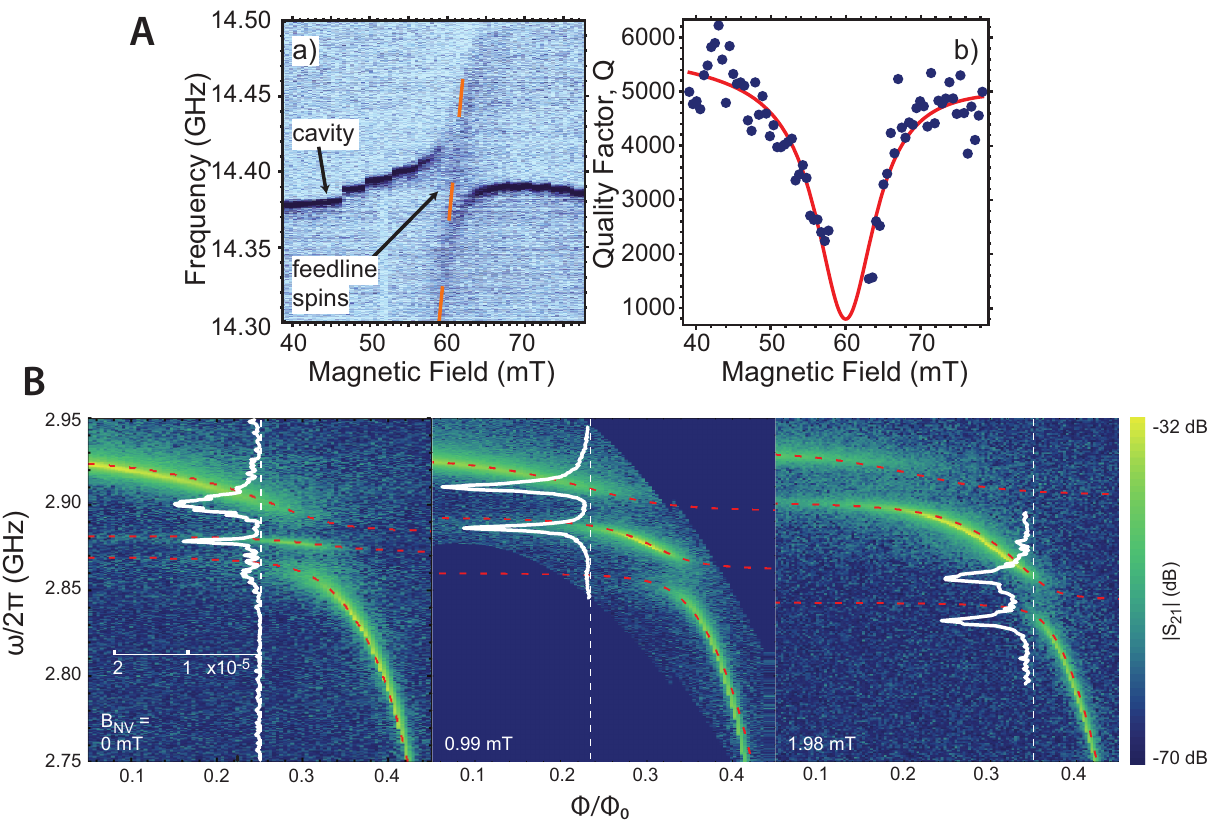}}
\caption{(A) The transmission spectrum of a superconducting stripline microwave resonator, fabricated on a ruby substrate, as a function of magnetic field. A coupling strength of 38~MHz is observed between the resonator and $\sim10^{12}$ Cr$^{3+}$ spins ($S=3/2$) located in the substrate, within the mode volume of the resonator. (B) The transmission $|S_{21}|$ of a resonator with a diamond on top shows two anti-crossing arising from the two $\Delta m_s=1$ transitions of the NV$^-$ centre, whose frequencies shift in opposite directions under a weak applied magnetic field $B_{\rm NV}$. The resonator itself incorporates a SQUID array, allowing frequency tuning through a locally  applied magnetic  flux $\Phi$. Panels (A) and (B) reprinted with permission from References~\cite{schuster10} and \cite{kubo10}, respectively, Copyright (2010) by the American Physical Society.}
\label{fig:superc}
\end{figure}

A key step in demonstrating a solid state microwave photon quantum memory based on electron spins is to demonstrate strong coupling between an electron spin ensemble and single microwave photons in a superconducting resonator. Typically, electron spins are placed in a magnetic field to achieve transition frequencies in the GHz regime, however, such fields can adversely affect the performance of superconducting resonators. One solution is to apply the magnetic field strictly parallel to the plane of thin film superconductors --- this has been used to observed coupling to a) the organic radical DPPH deposited over the surface of the resonator; b) paramagnetic Cr$^{3+}$ centres within the ruby substrate on which the resonator is fabricated; and c) substitutional nitrogen defects within a sample of diamond glued onto the top of the niobium resonator (see \Fig{fig:superc}a)~\cite{schuster10}. Another solution is to seek out electron spins which possess microwave transition frequencies in the absence of an applied magnetic field --- indeed this would appear the most attractive because the superconducting qubits themselves (which are generally made of aluminium) are unlikely to survive even modest magnetic fields. Thanks to their  zero-field splitting of $\sim3$~GHz, NV$^-$ centres in diamond have been used to observe strong coupling to a frequency-tuneable superconducting resonator in the absence of an appreciable magnetic field ($B<3$~mT)~\cite{kubo10} (see \Fig{fig:superc}b). The ability to rapidly tune the resonator into resonance with the spin ensemble should enable the observation of coherent oscillations between the two, and ultimately the storage of a single microwave photon in the spin ensemble.

An alternative approach is to couple an electron spin, or spins, directly to the syperconducting qubit, as proposed by Marcos et al.~\cite{marcos10}, or similarly to boost the coupling of a superconducting resonator to a single spin through the use of a persistent current loop~\cite{twamley10}.
Having stored the superconducting qubit state within a collective state of electron spins, it is then possible to transfer the state into a collective state of nuclear spins~\cite{wu09}, using the coherence transfer methods described in \Sec{sec:nucspins}, thus benefitting from the even longer nuclear coherence times.\\
\\
\begin{tabular}{|p{13cm}|}
\hline
{\bf Single-spin electron spin resonance}\\
Some of the approaches described here could be extended to enable general single-spin ESR.\\
{\bf Cavities} - The greatly reduced mode volume and high Q-factor of nanoscale superconducting resonators should lead to sufficient coupling between a single spin and the microwave field to permit single spin ESR. 
Typical values of $Q \sim10^5$ and a cavity of volume 1~pL~(1~$\mu$m$^3$) and frequency 1~GHz  would suggest a single spin will emit a scattered microwave photon every 10~ms~\cite{schuster10}. \\
{\bf Optics} - Single NV$^-$ electron spins can be measured close to the diamond surface and are sensitive to long-range dipolar coupling~\cite{neumann10},  enabling the indirect detection of other electron spins~\cite{maze08,balasubramanian08}. A nanocrystal of diamond could be mounted on a probe tip which scans over a surface, or the spins of interest could be deposited on a suitable diamond substrate.\\
{\bf Charge} -  Transport measurements of a carbon nanotube-based double QD device have shown the signature of coupling between a QD and a nearby electron spin~\cite{chorley10}. Carbon nanotubes can be controllably activated and functionalised with other molecular species~\cite{rodriguez09}. Combining such structures with magnetic resonance techniques could permit single spin detection of an attached electron spin~\cite{wabnig09, giavaras10}.\\
\hline
\end{tabular}

\section{Summary and outlook}

In this review, we have attempted in this review to systematically look at the coupling of electron spin to other degrees of freedom capable of hosting quantum information: nuclear spin, charge, optical photons, and superconducting qubits. We have not been able to cover all possibilities. For example, we have not mentioned 
electron spins coupled to mechanical states, which are explored in a recent proposal~\cite{rabl10}. 

In many cases, substantial benefit arises from bringing together more than one of the hybridising schemes discussed in this article. For example, in NV$^-$ centres, one can combine three degrees of freedom: the optical electronic transition, the electron spin, and the nuclear spin of a nearby \cthirteen. Alternatively, a hybrid electrical and optical method for measuring single electron or nuclear spin states of P-donors in silicon has been demonstrated, which exploits the ability to selectively optically excite a bound exciton state of the P-donor and a nearby QPC~\cite{sleiter10}.

Nevertheless, it is clear that electron spins play an essential role in a wide range of proposals for hybridising quantum information in the solid state. This can be attributed to their balance of portability, long-lived coherent behaviour, and versatility in coupling to many varied degrees of freedom.  

\subsection{Summary Points}
\begin{enumerate}
\item A versatile qubit can be represented in the electron spin of quantum dots, impurities in solids, organic molecules, and free electrons in the solid state.
\item A coupled electron spin can provide many advantages to a nuclear spin qubit, including high fidelity initialisation, manipulation on the nanosecond timescale, and more sensitive measurement.
\item The state of an electron spin can be coherently stored in a coupled nuclear spin, offering coherence times exceeding seconds ({\bf \Fig{fig:qmemory}}).
\item The interaction of an electron spin with many optical photons can be used for single spin measurement ({\bf \Fig{atature1fig}}), and manipulation on the picosecond timescale, while coupling to a single photon could be used to generate entanglement between two macroscopically separated electron spins ({\bf \Fig{mbqc}}). 
\item The electron spin of lithographically defined quantum dots can be measured through spin-dependent tunnelling between and off dots, and the use of a local quantum point contact.
\item A similar technique can been applied to the electron spin of a donor in silicon, the important difference being that the electron tunnels off the donor directly onto a single-electron transistor island, offering high fidelity single-shot measurement ({\bf \Fig{morello}}).
\item Electron spin could be used to store the state of a superconducting qubit. An important step towards achieving this has been demonstrated in the observation of strong coupling between an electron spin ensemble and a superconducting microwave stripline resonator ({\bf \Fig{fig:superc}}).
\end{enumerate}

\section{Annotations to References}
\begin{itemize}
\item \cite{morello10} Demonstration of single-shot single spin measurement in silicon, of what is most likely a P-donor
\item \cite{morton:qmemory} Demonstrates storage and retrieval of a coherent electron spin state in a P-donor nuclear spin, giving a quantum memory lifetime exceeding seconds.
\item \cite{schuster10}, \cite{kubo10} Offer parallel demonstrations of coupling of spin ensembles to superconducting microwave stripline resonators at milliKelvin temperatures
\item \cite{neumann10} Using opticla spin measurement, observes coupling between a single pair of NV$^-$ centres in diamond, separated by $\sim$100~\AA.
\item \cite{morton:bangbang} Show manipulation of the nuclear spin through microwave operations on the electron spin, and the use of this in dynamic decoupling
\item \cite{berezovsky08} Demonstrates optical initialization, manipulation and measurement of a single spin in a quantum dot.
\item\cite{vamivakas09} Demonstrates spin-dependent resonant fluorescence in a quantum dot.
\item \cite{benjamin09} Reviews measurement based quantum computing in solid state systems.
\item \cite{hanson07} Reviews the principles behind spin-to-charge conversion in quantum dots.
\end{itemize}




\end{document}